\documentclass[a4paper,11pt]{article}
\pdfoutput=1
\usepackage{pos}
\usepackage{comment}
\usepackage{lineno}
\usepackage{graphicx}
\usepackage{subcaption}
\title{Line emission search from DM annihilation in the Galactic Center with LST-1}
\author*[a]{Abhishek Abhishek}
\author[b]{Shotaro Abe}
\author[b, c, d]{Tomohiro Inada}
\author[a]{Sofia Ventura}
\author[e]{Michele Doro}
\author[b,f]{Masahiro Teshima}
\author[a]{Gaia Verna}
\onbehalf{on behalf of the CTAO-LST Project}
\affiliation[a]{University of Siena \& INFN Pisa, Italy}
\affiliation[b]{ICRR, UTokyo, Japan}
\affiliation[c]{Kyushu University, Japan}
\affiliation[d]{CERN, Switzerland}
\affiliation[e]{University of Padova, Italy}
\affiliation[f]{MPP, Germany}

\emailAdd{a.abhishek@student.unisi.it}
\abstract{Dark Matter (DM) remains a great mystery in modern physics. Among various candidates, the weakly interacting massive particles (WIMPs) scenario stands out and is under extensive study. The detection of the hypothetical gamma-ray emission from WIMP annihilation could act as a direct probe of electroweak-scale interactions, complementing DM collider searches and other direct DM detection techniques. At very high energies (VHE), WIMP self-annihilation is expected to produce gamma rays together with other Standard Model particles. The Galactic Center (GC), due to its relative proximity to the Earth and its high expected DM density, is a prime target for monoenergetic line searches. Imaging Atmospheric Cherenkov Telescopes (IACTs) have placed strong constraints on the DM properties at the GC, with the Major Atmospheric Gamma-ray Imaging Cherenkov (MAGIC) providing the most stringent limits from 20 TeV to 100 TeV exploiting Large Zenith Angle (LZA) observations. However, the limited field of view (FoV) of the MAGIC telescopes (< 3.5° ) prevented a detailed study of the extended region around the GC in which an enhanced DM density is expected.
The first Large-Sized Telescope (LST-1) of the Cherenkov Telescope Array Observatory (CTAO), located at the Roque de Los Muchachos Observatory (La Palma, Spain) close to the MAGIC site, has been observing the GC since 2021. With its wide FoV of 4.5°, LST-1 could contribute significantly to the WIMPs search at the GC. The observations are performed at LZA (ZA > 58°), which, while required due to the source's low altitude, also optimizes the detection of gamma rays up to 100 TeV and beyond. We present a study of the systematic uncertainties in WIMP line emission searches with LST-1. Our work examines the instrument response functions for LZA observations, background rejection in monoscopic mode, and includes updated results from simulations, highlighting new methods for spectral line searches.}

\ConferenceLogo{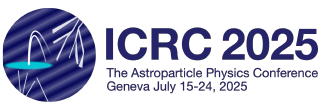}

\FullConference{39th International Cosmic Ray Conference (ICRC2025)\\
 15–24 July 2025\\
Geneva, Switzerland\\}

\begin{document}
\maketitle

\section{DM Line Signatures in Gamma-Ray Observations}

A wide range of astrophysical and cosmological evidence indicates that most of the Universe’s matter is non-baryonic and invisible, commonly referred to as dark matter (DM), with Weakly Interacting Massive Particles (WIMPs) being leading candidates~\cite{conrad_indirect_2014}. Indirect searches target the annihilation or decay products of DM such as gamma rays from regions of high DM density like the Galactic Center ~\cite{bergstrom_observability_1998}. Monochromatic gamma-ray lines from DM annihilation, such as $\chi\chi \rightarrow \gamma\gamma$ and $\chi\chi \rightarrow \gamma Z$, would appear as sharp spectral features near the DM mass and could provide 'smoking gun' signatures of WIMP DM ~\cite{foster_search_2023}. Among WIMP candidates, wino and higgsino SUSY particles are especially promising for indirect detection via gamma-ray lines in the VHE ($\geq$100 GeV) regime~\cite{hisano_explosive_2004}. Both ground-based Cherenkov telescopes and space-based instruments have placed strong constraints on these scenarios, with recent analyses reaching sensitivities in the TeV mass range near the canonical thermal relic cross-section~\cite{hess_collaboration_search_2018, magic_collaboration_search_2023, foster_search_2023}.
In this work, we present an overview of the DM line search at the Galactic Center, focusing on systematic studies with the first Large-Sized Telescope (LST-1) of CTAO for the search for line-like gamma-ray signals from DM annihilation.

\section{Indirect DM Searches Toward the Galactic Center with LSTs at Large Zenith Angles}
\label{GC_and_background}

The Galactic Center (GC) is a prime target for indirect DM searches due to its high predicted DM density ~\cite{abe_dark_2024}. Hence this is one of the prime location for such searches for the current ground-based IACTs and the upcoming Cherenkov Telescope Array (CTAO). The H.E.S.S. collaboration study of GC for DM line search in 2018 have produced quite stringent results among the IACTs starting from the energies of 300 GeV~\cite{hess_collaboration_search_2018}. LST-1 has been observing GC since 2021 and a study of the extended region of GC can be found in \textit{Shotaro Abe-ICRC2025}. The location of LST-1 site at the Roque de Los Muchachos, Canary Islands, makes the observation of the GC only possible at large zenith angles (LZA) of > $58^{\circ}$ (low altitudes), which comes with downsides such as higher energy thresholds (For, e.g., > 400 GeV for MAGIC GC spectral analysis ~\cite{acciari_magic_2020}) due to dimmer shower images and increased background rates at low altitudes, but it also has some upsides like the greater effective collection area, thereby boosting sensitivity for VHE gamma ray showers ( above $\sim$ 1 TeV), which is crucial for probing DM particles with masses in the TeV range~\cite{abe_dark_2024, abe_discovering_2025}. The MAGIC telescopes, which are also located at the same site as LST-1, hence bounded by the same limitations, had performed a study of DM lines from GC using over 200 hours of data, producing best limits above 20 TeV energies among the IACTs ~\cite{magic_collaboration_search_2023}. LST-1 is a well-suited telescope to study DM at multi-TeV mass range given it's wide ($4.5^{\circ}$) FoV. 

The analysis models dark matter line signals alongside background components to detect or constrain DM parameters. Accurate background modeling is crucial to distinguish potential DM signals from astrophysical and cosmic-ray backgrounds, enabling robust detection or upper limits. For line search, we replace spatial background subtraction with the sliding window technique~\cite{vertongen_hunting_2011}, fitting energy spectra in narrow windows around candidate DM line energies. Within each window, we model background locally as a power law and search for DM lines (convolved with energy resolution) above this baseline. Systematic uncertainties from local power-law approximations are evaluated using OFF-source data, while telescope/instrument response functions systematics will be covered in the future work.

\section{Data Analysis}

In this work, we address systematics from background modeling as described in \ref{GC_and_background}. Let's look at low level and high level analysis of the OFF (background) data we have from LST-1. 

\vspace{0.5cm}

\noindent \textbf{OFF source data analysis}

\subsection{Low-level analysis}
\label{low-level}
In IACT data analysis, low-level analysis refers to the processing of raw data taken by the telescope, which here is performed using \textit{lstchain} package ~\cite{lstchain_adass_2020}. For this analysis, we utilize $\sim$ 6.8 hours of OFF data of the transient sources observed at large zenith angles (>45°) during period April 2021 to September 2024, where target sources showed no significant emission. For this analysis, OFF-source data were chosen to match the observing conditions of the Galactic Center as closely as possible, specifically in terms of zenith angle and dark night conditions. The event selection criteria is carefully optimized for gammaness cuts. The gammaness\footnote{It is a \textit{score} used in IACT analysis indicating how likely the event is a gamma as compared to a proton or other cosmic-ray particle} cut was optimized for the analysis by studying the quality factor (Q-factor) metric [Sec:\ref{results}]. The data were processed using standard selection criteria with optimized gammaness cuts and interpolated instrument response functions. This low-level analysis yields data products containing all the information necessary for subsequent scientific interpretation. 

\subsection{High-level analysis}
\label{high-level-analysis}
\begin{figure}[h]
    \centering
    \includegraphics[width=0.5\linewidth]{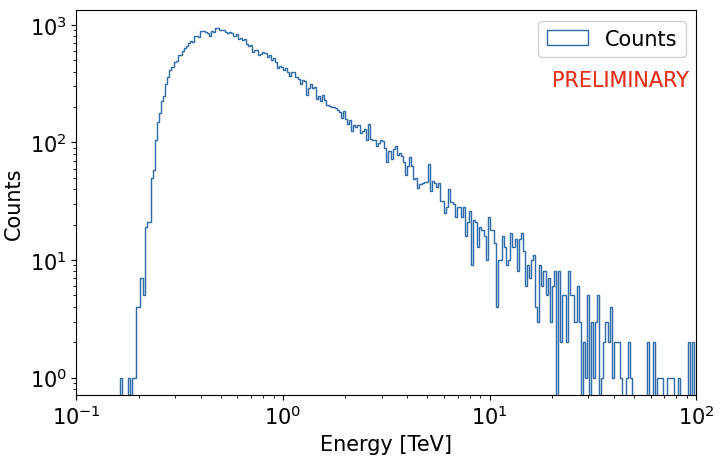}
    \caption{Energy distribution of background counts in the studied OFF regions}
    \label{fig:bkg_off}
\end{figure}

High-level analysis refers to scientific interpretation from modeling of processed data. This is performed using \textit{Gammapy} package \cite{acero_gammapy_2025}. Using the DL3 files produced in [Sec: \ref{low-level}], we produce Gammapy objects: MapDataset (3-dimensional dataset) by sampling the counts in a narrow, logarithmically spaced energy axis (100 bins per decade) in order to constrain the sharp line feature convolved with the energy resolution, spread across a few tens of bins. These MapDataset of the OFF-source regions were converted to SpectrumDataset (1-dimensional dataset) by summing over the spatial axes within the region of interest (ROI). For this analysis, we adopt a region of interest (ROI) defined as a $2.1^\circ$ circular area centered on the source, which efficiently utilizes the $2.5^\circ$ of the field of view as the observations are conducted at a $0.4^\circ$ offset. The resulting count spectra were then stacked producing the composite OFF spectrum shown in Fig:~\ref{fig:bkg_off}

Systematic uncertainties from linear background modeling are evaluated by comparing power-law (PL) and log-parabola (LP) fits, with two and three free parameters respectively, while spectral features mimicking/masking lines are also assessed. The sliding window width is defined as $[E/(1+\sigma_{E})^{\mu}, E(1+\sigma_{E})^{\mu}]$, where $\sigma_{E}$ is the energy resolution at energy $E$, and $E$ is the DM mass adjusted for the typical difference between true and reconstructed energy (energy bias), as defined in the IRF, and $\mu$ scales the width (optimization details in Sec.~\ref{results}). Following MAGIC-2023's $\pm 4\sigma_{\mathrm{E}}$ window~\cite{magic_collaboration_search_2023}, we explore similar widths. The DM line signal is modeled using flux for Cirelli et al.'s $\gamma\gamma$ spectrum~\cite{cirelli_pppc_2011} in \textit{Gammapy}, multiplied by a \textit{scale} parameter. The background PL model with We perform a likelihood scan of \textit{scale}, multiplying the best-fit value by the thermal relic cross-section to estimate the upper limit on $\langle\sigma v\rangle$. Expected sensitivity is computed using the Asimov procedure ~\cite{cowan_asymptotic_2011}.

\section{Results}
\label{results}

\begin{figure}[h]
    \centering
    \includegraphics[width=0.5\linewidth]{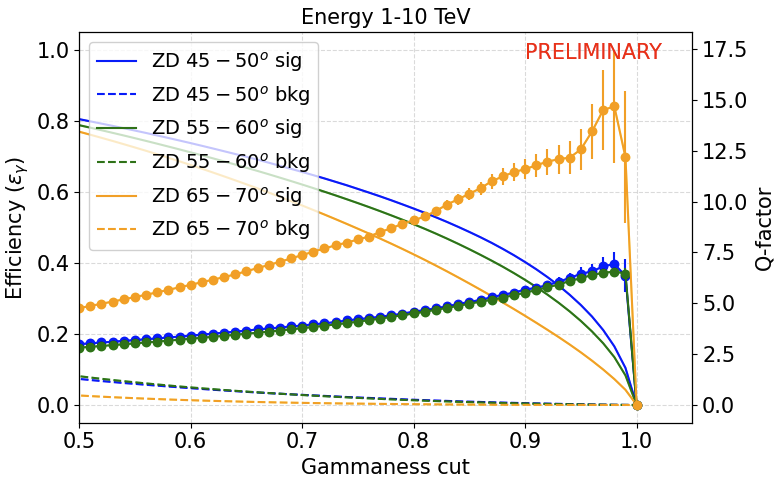}
    \caption{The gammaness efficiency (solid lines) and background efficiency (dashed lines) are plotted for gammaness cuts, and the corresponding Q-factor values (solid line with circle markers) are shown in the y-axis on right.}
    \label{fig:q-factor}
\end{figure}

The optimization of event cut based on the differentiability of a signal event (gamma) from a background event is performed using the Q-factor metric : $\mathrm{Q} = \mathrm{\epsilon_{sig}}/\mathrm{\sqrt{\epsilon_{bkg}}}$ where $\mathrm{\epsilon_{sig}}$ and $\mathrm{\epsilon_{bkg}}$ represent the survival fractions of signal ($\gamma_{\mathrm{MC}}$) and background (Real events dominated by hadron background events) events post-cut, respectively. In the case of OFF-source data, and using the MC event files produced for the respective declination lines, we have evaluated the Q-factor for different cuts as shown in the Fig:\ref{fig:q-factor}

Though the Q-factor peaks near a strict gammaness efficiency ($\epsilon_{\gamma}$) cut of $\sim$20\% for upto $65^{\circ}$ zd. The gammaness cut corresponding to 20\% $\epsilon_{\gamma}$ is very strict compared to standard LST analysis cut of 70\% \cite{abe_observations_2023}. Using this 20\% $\epsilon_{\gamma}$ cut as opposed to standard, we gain  sensitivity of 50\% for the ZD range $45^{\circ}-$$60^{\circ}$ and $\sim 80\%$ for higher ZD. We verified the usage of such a strict cut by performing a cross-check with LZA observations of Crab Nebula, the standard candle for gamma ray astronomy. Even using an energy dependent $\epsilon_{\gamma}$ cut of 20\%,  we produced Crab Nebula flux points at LZA within 25\% uncertainties from the LST-1 flux points at nominal zenith angles \cite{abe_observations_2023} in the energy range of 900 GeV to 15 TeV. The resulting energy resolution at $3$ TeV for $50^\circ, 60^{\circ}$ and $70^{\circ}$ ZD, after applying the optimized gammaness cut, are 18\%, 22\% and 28\% respectively. 

\begin{figure}[htp]
    \centering
    \begin{subfigure}[t]{0.46\textwidth}
        \centering
        \includegraphics[width=\linewidth]{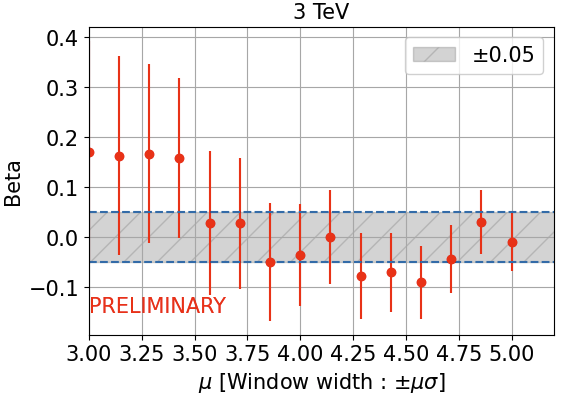}
        \caption{}
        \label{fig:sub1}
    \end{subfigure}
    \hspace*{\fill}
    \begin{subfigure}[t]{0.50\textwidth}
        \centering
        \includegraphics[width=\linewidth]{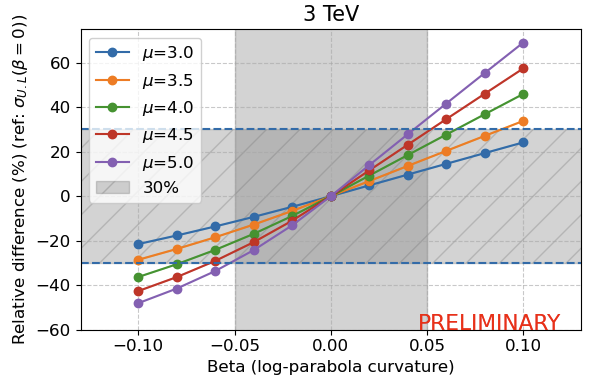}
        \caption{}
        \label{fig:sub2}
    \end{subfigure}
    \caption{\textit{left}: $\beta$ is varying as the window width changes showing the non-linear small scale features on the background spectrum. \textit{right}: The relative sensitivity difference of statistical and systematic values displaying changing $\beta$ hindering the computation of DM parameter}
    \label{fig:both}
\end{figure}

However the OFF source data was produced using a global gammaness cut corresponding to 20\% $\epsilon_{\gamma}$ to ensure smooth counts distribution across energy binning of IRFs. This resulting data is analyzed to study the systematics associated with background modeling. This study is focused on search of a line from DM mass of 3 TeV. The sliding window technique assumes a linear counts distribution within the window, the uncertainty of this assumption can be seen by fitting a non-linear model, for example, LP: \[\Phi(E) = \Phi_{0} (\frac{E}{E_{0}})^{-\alpha - \beta \times log (\frac{E}{E_0})}\] with a $\beta$ parameter adding curvature to the assumed linear PL model and $E_{o}$ as 3 TeV. 
\noindent When the spectrum around 3 TeV is fitted with the LP model for various window width ($\mu$ = [3.5, 5]) we got the $|\beta|$ is < 0.05 as shown in the Fig:\ref{fig:sub1} and the $\Delta\chi^{2}_{\mathrm{PL}-\mathrm{LP}}$ < 1 for $\pm 4\sigma_{\mathrm{E}}$ indicates that LP model does not provide a significantly better fit than PL at the 95\% confidence level, for all the window width of $3.5 < \mu < 5$. 

The systematic floor for the DM cross-section sensitivity arises from background model mismatches: when fitting a PL background to data with intrinsic curvature ($\beta \neq 0$), we observe a bias in the computed upper limits for the DM annihilation cross-section. For $\beta$ within $\pm 0.05$, the sensitivity degrades by up to 30\% relative to the ideal $\beta=0$ case (Fig.~\ref{fig:sub2}). This defines a systematic floor of $\sim$10\%, 20\%, and 30\% for 3 TeV DM lines using $\pm3$, $\pm4$, and $\pm5,\sigma_E$ windows, respectively.

These results motivate a more detailed study of systematics and support the use of an energy-dependent window width to optimize sensitivity in dark matter line searches.

%\clearpage
\section{Summary}
In this contribution, we have presented the study of background modeling from the counts spectrum of the OFF source regions at similar conditions of Galactic Center with the LST-1 telescope. Our analysis highlights the importance of careful analysis of background before achieving sensitivity for line-like gamma-ray signatures in the TeV mass range. We have shown the optimization of the analysis for large zenith angle observations using quality-factor and rigorously quantifying the background-related systematics for energy window. This work details the methodology and systematics for the DM line search with LST-1, with selected preliminary results to be shown in the accompanying presentation. Optimization and implementation to the GC data, defining an energy dependent window width while spanning over the energy range for DM line search, selection of optimal ROI to utilize LST-1's wide FoV, and improving the systematic uncertainty floor with better background suppression using the upcoming 4 LSTs in CTAO-North site are the future prospects of this study.

\acknowledgments 
\tiny{
We gratefully acknowledge financial support from the following agencies and organisations:
Conselho Nacional de Desenvolvimento Cient\'{\i}fico e Tecnol\'{o}gico (CNPq), Funda\c{c}\~{a}o de Amparo \`{a} Pesquisa do Estado do Rio de Janeiro (FAPERJ), Funda\c{c}\~{a}o de Amparo \`{a} Pesquisa do Estado de S\~{a}o Paulo (FAPESP), Funda\c{c}\~{a}o de Apoio \`{a} Ci\^encia, Tecnologia e Inova\c{c}\~{a}o do Paran\'a - Funda\c{c}\~{a}o Arauc\'aria, Ministry of Science, Technology, Innovations and Communications (MCTIC), Brasil;
Ministry of Education and Science, National RI Roadmap Project DO1-153/28.08.2018, Bulgaria;
Croatian Science Foundation (HrZZ) Project IP-2022-10-4595, Rudjer Boskovic Institute, University of Osijek, University of Rijeka, University of Split, Faculty of Electrical Engineering, Mechanical Engineering and Naval Architecture, University of Zagreb, Faculty of Electrical Engineering and Computing, Croatia;
Ministry of Education, Youth and Sports, MEYS  LM2023047, EU/MEYS CZ.02.1.01/0.0/0.0/16\_013/0001403, CZ.02.1.01/0.0/0.0/18\_046/0016007, CZ.02.1.01/0.0/0.0/16\_019/0000754, CZ.02.01.01/00/22\_008/0004632 and CZ.02.01.01/00/23\_015/0008197 Czech Republic;
CNRS-IN2P3, the French Programme d’investissements d’avenir and the Enigmass Labex, 
This work has been done thanks to the facilities offered by the Univ. Savoie Mont Blanc - CNRS/IN2P3 MUST computing center, France;
Max Planck Society, German Bundesministerium f{\"u}r Bildung und Forschung (Verbundforschung / ErUM), Deutsche Forschungsgemeinschaft (SFBs 876 and 1491), Germany;
Istituto Nazionale di Astrofisica (INAF), Istituto Nazionale di Fisica Nucleare (INFN), Italian Ministry for University and Research (MUR), and the financial support from the European Union -- Next Generation EU under the project IR0000012 - CTA+ (CUP C53C22000430006), announcement N.3264 on 28/12/2021: ``Rafforzamento e creazione di IR nell’ambito del Piano Nazionale di Ripresa e Resilienza (PNRR)'';
ICRR, University of Tokyo, JSPS, MEXT, Japan;
JST SPRING - JPMJSP2108;
Narodowe Centrum Nauki, grant number 2023/50/A/ST9/00254, Poland;
The Spanish groups acknowledge the Spanish Ministry of Science and Innovation and the Spanish Research State Agency (AEI) through the government budget lines
PGE2022/28.06.000X.711.04,
28.06.000X.411.01 and 28.06.000X.711.04 of PGE 2023, 2024 and 2025,
and grants PID2019-104114RB-C31,  PID2019-107847RB-C44, PID2019-104114RB-C32, PID2019-105510GB-C31, PID2019-104114RB-C33, PID2019-107847RB-C43, PID2019-107847RB-C42, PID2019-107988GB-C22, PID2021-124581OB-I00, PID2021-125331NB-I00, PID2022-136828NB-C41, PID2022-137810NB-C22, PID2022-138172NB-C41, PID2022-138172NB-C42, PID2022-138172NB-C43, PID2022-139117NB-C41, PID2022-139117NB-C42, PID2022-139117NB-C43, PID2022-139117NB-C44, PID2022-136828NB-C42, PDC2023-145839-I00 funded by the Spanish MCIN/AEI/10.13039/501100011033 and “and by ERDF/EU and NextGenerationEU PRTR; the "Centro de Excelencia Severo Ochoa" program through grants no. CEX2019-000920-S, CEX2020-001007-S, CEX2021-001131-S; the "Unidad de Excelencia Mar\'ia de Maeztu" program through grants no. CEX2019-000918-M, CEX2020-001058-M; the "Ram\'on y Cajal" program through grants RYC2021-032991-I  funded by MICIN/AEI/10.13039/501100011033 and the European Union “NextGenerationEU”/PRTR and RYC2020-028639-I; the "Juan de la Cierva-Incorporaci\'on" program through grant no. IJC2019-040315-I and "Juan de la Cierva-formaci\'on"' through grant JDC2022-049705-I. They also acknowledge the "Atracci\'on de Talento" program of Comunidad de Madrid through grant no. 2019-T2/TIC-12900; the project "Tecnolog\'ias avanzadas para la exploraci\'on del universo y sus componentes" (PR47/21 TAU), funded by Comunidad de Madrid, by the Recovery, Transformation and Resilience Plan from the Spanish State, and by NextGenerationEU from the European Union through the Recovery and Resilience Facility; “MAD4SPACE: Desarrollo de tecnolog\'ias habilitadoras para estudios del espacio en la Comunidad de Madrid" (TEC-2024/TEC-182) project funded by Comunidad de Madrid; the La Caixa Banking Foundation, grant no. LCF/BQ/PI21/11830030; Junta de Andaluc\'ia under Plan Complementario de I+D+I (Ref. AST22\_0001) and Plan Andaluz de Investigaci\'on, Desarrollo e Innovaci\'on as research group FQM-322; Project ref. AST22\_00001\_9 with funding from NextGenerationEU funds; the “Ministerio de Ciencia, Innovaci\'on y Universidades”  and its “Plan de Recuperaci\'on, Transformaci\'on y Resiliencia”; “Consejer\'ia de Universidad, Investigaci\'on e Innovaci\'on” of the regional government of Andaluc\'ia and “Consejo Superior de Investigaciones Cient\'ificas”, Grant CNS2023-144504 funded by MICIU/AEI/10.13039/501100011033 and by the European Union NextGenerationEU/PRTR,  the European Union's Recovery and Resilience Facility-Next Generation, in the framework of the General Invitation of the Spanish Government’s public business entity Red.es to participate in talent attraction and retention programmes within Investment 4 of Component 19 of the Recovery, Transformation and Resilience Plan; Junta de Andaluc\'{\i}a under Plan Complementario de I+D+I (Ref. AST22\_00001), Plan Andaluz de Investigaci\'on, Desarrollo e Innovación (Ref. FQM-322). ``Programa Operativo de Crecimiento Inteligente" FEDER 2014-2020 (Ref.~ESFRI-2017-IAC-12), Ministerio de Ciencia e Innovaci\'on, 15\% co-financed by Consejer\'ia de Econom\'ia, Industria, Comercio y Conocimiento del Gobierno de Canarias; the "CERCA" program and the grants 2021SGR00426 and 2021SGR00679, all funded by the Generalitat de Catalunya; and the European Union's NextGenerationEU (PRTR-C17.I1). This research used the computing and storage resources provided by the Port d’Informaci\'o Cient\'ifica (PIC) data center.
State Secretariat for Education, Research and Innovation (SERI) and Swiss National Science Foundation (SNSF), Switzerland;
The research leading to these results has received funding from the European Union's Seventh Framework Programme (FP7/2007-2013) under grant agreements No~262053 and No~317446;
This project is receiving funding from the European Union's Horizon 2020 research and innovation programs under agreement No~676134;
ESCAPE - The European Science Cluster of Astronomy \& Particle Physics ESFRI Research Infrastructures has received funding from the European Union’s Horizon 2020 research and innovation programme under Grant Agreement no. 824064.}

\noindent \textbf{Full Author List: CTAO-LST Project}

% POS format
% Generated on 2025-05-28
\tiny{\noindent
K.~Abe$^{1}$,
S.~Abe$^{2}$,
A.~Abhishek$^{3}$,
F.~Acero$^{4,5}$,
A.~Aguasca-Cabot$^{6}$,
I.~Agudo$^{7}$,
C.~Alispach$^{8}$,
D.~Ambrosino$^{9}$,
F.~Ambrosino$^{10}$,
L.~A.~Antonelli$^{10}$
% ...
The CTAO-LST Project
}\\

\tiny{\noindent$^{1}${Department of Physics, Tokai University, 4-1-1, Kita-Kaname, Hiratsuka, Kanagawa 259-1292, Japan}.
$^{2}${Institute for Cosmic Ray Research, University of Tokyo, 5-1-5, Kashiwa-no-ha, Kashiwa, Chiba 277-8582, Japan}.
$^{3}${INFN and Università degli Studi di Siena, Dipartimento di Scienze Fisiche, della Terra e dell'Ambiente (DSFTA), Sezione di Fisica, Via Roma 56, 53100 Siena, Italy}.
$^{4}${Université Paris-Saclay, Université Paris Cité, CEA, CNRS, AIM, F-91191 Gif-sur-Yvette Cedex, France}.
$^{5}${FSLAC IRL 2009, CNRS/IAC, La Laguna, Tenerife, Spain}.
$^{6}${Departament de Física Quàntica i Astrofísica, Institut de Ciències del Cosmos, Universitat de Barcelona, IEEC-UB, Martí i Franquès, 1, 08028, Barcelona, Spain}.
$^{7}${Instituto de Astrofísica de Andalucía-CSIC, Glorieta de la Astronomía s/n, 18008, Granada, Spain}.
$^{8}${Department of Astronomy, University of Geneva, Chemin d'Ecogia 16, CH-1290 Versoix, Switzerland}.
$^{9}${INFN Sezione di Napoli, Via Cintia, ed. G, 80126 Napoli, Italy}.
$^{10}${INAF - Osservatorio Astronomico di Roma, Via di Frascati 33, 00040, Monteporzio Catone, Italy}.
% ...
}

\end{document}